\documentclass[prl,twocolumn,aps,letterpaper, preprintnumbers,superscriptaddress]{revtex4}
\usepackage{upgreek}
\usepackage{hyperref}
\usepackage{verbatim}
\usepackage{amsmath}
\usepackage{latexsym}
\usepackage{revsymb}
\usepackage{yfonts}
\usepackage{ifthen}
\usepackage{tcolorbox}
\usepackage{natbib}
\usepackage{amsfonts}
\usepackage{amsmath}
\usepackage{amssymb}
\usepackage{amsthm}
\usepackage{graphicx}
\usepackage{bm}
\usepackage{bbm}
\usepackage{epsfig,color,amssymb}
\usepackage{subfigure}
\usepackage{amsfonts}
\usepackage{amscd}
\usepackage{amsmath}
\usepackage{multirow}
\usepackage{chemarrow}
\usepackage{dcolumn}
\usepackage{bm}
\usepackage{graphicx}
\usepackage{enumerate}
\usepackage{epsfig}
\usepackage{subfigure}
\usepackage{xcolor}
\usepackage{multirow}
\usepackage{ulem}
\usepackage{braket}
\usepackage{comment}
\usepackage{enumitem}
\usepackage{amsthm}
\usepackage{diagbox}

\renewcommand{\emph}[1]{\textit{#1}}

\usepackage{siunitx}

\newcommand{\ba}{\begin{array}}
\newcommand{\ea}{\end{array}}
\newcommand{\be}{\begin{equation}}
\newcommand{\ee}{\end{equation}}
\newcommand{\bea}{\begin{eqnarray}}
\newcommand{\eea}{\end{eqnarray}}

\newcommand{\bfe}{{\bf e}}
\newcommand{\bfeR}{{\bf e}_{\rm R}}
\newcommand{\bfeL}{{\bf e}_{\rm L}}

\newcommand{\bfr}{{\bf r}}

\newcommand{\bfE}{{\bf E}}
\newcommand{\bfH}{{\bf H}}

\begin{document}
%\title{Topological photonic quasiparticles in gradient-index microlenses for ultra-capacity secure encryption}
%\title{Topological multiskyrmionic quasiparticles in photonic gradient-index lenses}
\title{Photonic torons, topological phase transition and tunable spin monopoles}
%\title{Reconfigurable topological poly-merons and poly-skyrmions in photonic gradient-index media}
\author{Haijun Wu}
\affiliation{Wang Da-Heng Center, Heilongjiang Key Laboratory of Quantum Control, Harbin University of Science and Technology, Harbin 150080, China}
\affiliation{Centre for Disruptive Photonic Technologies, School of Physical and Mathematical Sciences, Nanyang Technological University, Singapore 637371, Singapore}
\affiliation{Department of Electrical and Computer Engineering, National University of Singapore, Singapore, 117583, Singapore}
\author{Nilo Mata-Cervera}
\affiliation{Centre for Disruptive Photonic Technologies, School of Physical and Mathematical Sciences, Nanyang Technological University, Singapore 637371, Singapore}
\author{Haiwen Wang}
\affiliation{Department of Applied Physics, Stanford University, Stanford, California 94305, USA}
\author{Zhihan Zhu}
\affiliation{Wang Da-Heng Center, Heilongjiang Key Laboratory of Quantum Control, Harbin University of Science and Technology, Harbin 150080, China}
\author{Cheng-Wei Qiu}
\affiliation{Department of Electrical and Computer Engineering, National University of Singapore, Singapore, 117583, Singapore}
%\author{Shanhui Fan}
%\affiliation{Department of Electrical Engineering, Stanford University, Stanford, California 94305, USA}
\author{Yijie Shen}\email{yijie.shen@ntu.edu.sg}
\affiliation{Centre for Disruptive Photonic Technologies, School of Physical and Mathematical Sciences, Nanyang Technological University, Singapore 637371, Singapore}
\affiliation{School of Electrical and Electronic Engineering, Nanyang Technological University, Singapore 639798, Singapore}

\date{\today}

%%%%%%%%%%%%%%%%%%%%%%%%%%%%%%%%%%%%%%%%%%

\begin{abstract}
\noindent Creation and control of topological complex excitations play crucial roles in both fundamental physics and modern information science. Torons are a sophisticated class of 3D chiral polar topological structures with both skyrmionic quasiparticle textures and monopole point defects, so far only observed in liquid crystal nonpolar models. Here, we experimentally construct torons with the photonic spin of vector structured light and demonstrate the topological phase transitions among diverse topological states: torons, hopfions, skyrmioniums and monopole pairs. We can also continually tune the toron’s chirality and the helical spin textures of emerging monopole pairs. The birth of photonic torons and tunable monopoles opens a flexible platform for studying nontrivial light-matter interaction and topological informatics. 
\end{abstract}

\maketitle

\noindent Topological structures, including defects located at 0D point and 1D line defects and textures filled in 2D, 3D and higher dimensions, and the phase transitions among which play important roles in modern physics~\cite{lin2023topological,haldane2017nobel,bernevig2022progress}. Manipulation of topological complex excitations in electromagnetic fields and condensed matter promises advanced high-density and robust data encoding, storage and transfer~\cite{han2022high,chen2024all,fert2017magnetic,papasimakis2016electromagnetic,wan2023ultra}. 
As a typical topological 0D electromagnetic defect, magnetic monopole is still to be found despite the symmetry of Maxwell's equations, but plays an important role in modern physics due to the recent studies of its counterpart in synthetic fields or materials~\cite{carrigan1983magnetic,castelnovo2008magnetic,qi2009inducing,beche2014magnetic,kanazawa2020direct,rana2023three,marques2024magneto}. 
%\HW{(I am not sure you can say theoretically magnetic monopoles don't exist. I would just say they are not found despite the symmetry of Maxwell's equations.)} 
Although it is impossible to be directly generated in electromagnetic waves, the monopoles were recently theoretically modeled in optical spins~\cite{wang2022topological,wang2023photonic}. 
In addition to the 0D and 1D defects, 2D and 3D topological quasiparticles, such as skyrmions~\cite{tsesses2018optical, du2019deep, davis2020ultrafast,bogdanov2020physical,foster2019two}, 
merons~\cite{dai2020plasmonic,yu2018transformation,jani2021antiferromagnetic}, toroids~\cite{shen2021supertoroidal,zdagkas2021observation,wan2022toroidal,shen2024nondiffracting}, hopfions~\cite{zheng2023hopfion,shen2023topological,rybakov2022magnetic,sugic2021particle}, and other 3D chiral knotted structures~\cite{tai2019three,tang2021magnetic,ackerman2017static}, are constructed in localized continue vector fields, which have recently garnered significant attention in magnetic materials~\cite{gobel2021beyond}, soft matter~\cite{wu2022hopfions}, and photonics~\cite{shen2024optical}.

Torons, as a novel class of topological excitations, hold both features of point defects and quasiparticle textures, while the exact polar structure of which has never been observed in any physical system, only the nonpolar version of which was proposed in liquid crystals interacted with structured light~\cite{smalyukh2010three}. Since the first observation, torons showed their powerful topological transformation and phase transitions among diversified states including skyrmion, hopfion, and M\"obius stripes~\cite{ackerman2017diversity,tai2020surface,peixoto2024mechanical,zhao2023liquid}. Torons can also be clusterized into topological macromolecules with solitonic analogues of polymeric materials~\cite{zhao2023topological}. In contrast to other static topological excitations, torons recently showed unique dynamics, for instance, their co-assembly when interacting with optical solitons~\cite{poy2022interaction}, and chiral texture evolution of monopoles driven by external electromagnetic fields~\cite{tai2024field}. However, all the existing torons were studied as nonpolar scheme due to the head-tail symmetry of liquid crystals. The exact polar torons have been theoretically studied in magnets~\cite{li2022mutual,liu2018binding}, but never been observed in any physical system other than liquid crystals.

\begin{figure*}
\centering
  \includegraphics[width=0.95\linewidth]{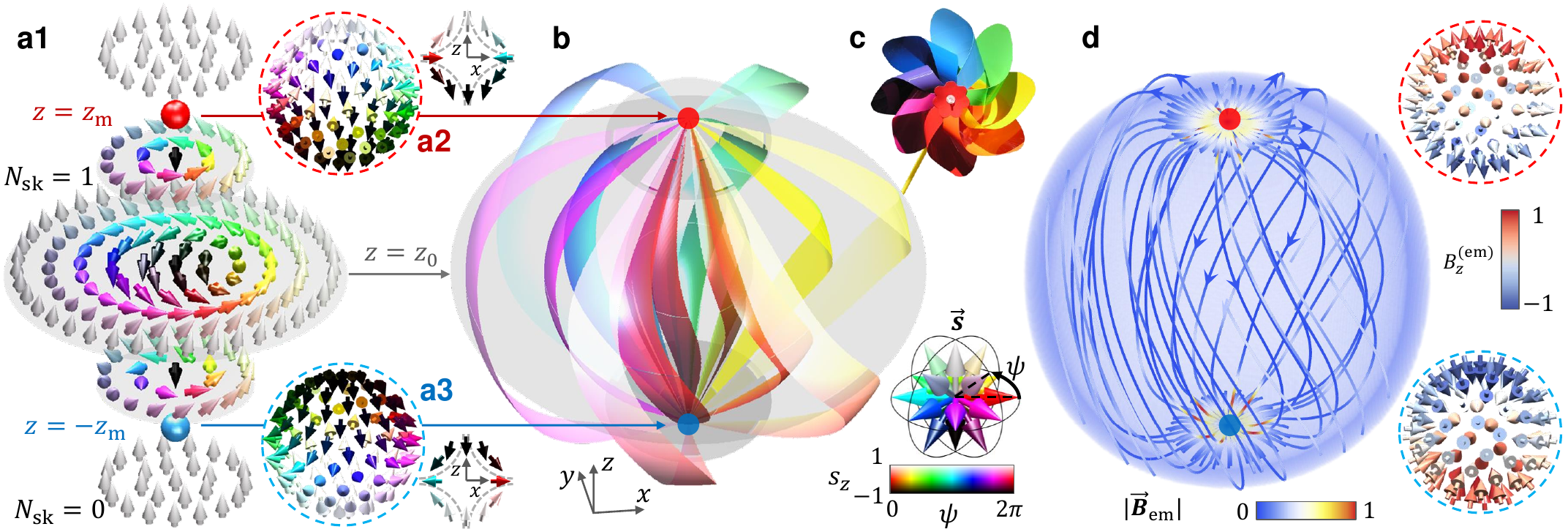}
  \caption{\textbf{Concept of toron topology.} \textbf{a1}, 3D spin distribution with subspace 2D textures highlighted in selective transverse cross-sections, where the structures of two spin monopole defects at $z=\pm z_\text{m}$ are highlighted by red and blue in \textbf{a2} and \textbf{a3}. \textbf{b}, Corresponding assembly of isospin fiber bundles obeying the spin density vector $\bf S$ of the toron. As indicated by the color map, the fiber bundle state is mapped to the HSL color space, where lightness encodes $s_z$, and the spin orientation $\psi=\text{Arg}[s_x+is_y]$ is represented by the hue, resulting in a rainbow-like variation across the isospin line cluster. Ideally, the isospin surfaces form closed surfaces for each value of $s_z$. However, to enhance visualization, specific spin orientations are selectively removed for different $s_z$ values, producing a multi-layered pinwheel-like structure of the toron bundle, \textbf{c}, like the structure of a pinwheel. For simplicity, in subsequent figures, only the monolayer of the isospin line cluster for $s_z=0$ is illustrated for representation of toron. \textbf{d}, The emergent magnetic field ${\bf B}_\text{em}$, derived from the spin density vector $\bf S$ of the toron, is represented by isosurfaces of its scalar magnitude and streamlines connecting the top (source-type) and bottom (sink-type) monopoles.}
  \label{f1}
\end{figure*}

%In electromagnetic fields, it was impossible to directly construct a toron or monopole due to the violation of Maxwell's equations.  While, 
Here, we present torons and monopoles in free-space electromagnetic fields, the vector textures of which are constructed by the electromagnetic spin angular momentum (SAM) or optical spin for short. Applying vector structured light modulation technologies~\cite{shen2021rays,forbes2021structured,he2022towards}, we successfully observe, for the first time, optical spin torons and realize controlled phase transitions among skyrmion, hopfion, and monopole pair states. We also demonstrate flexibly tunable toron chirality and helical textures of spin monopoles. Note that all the topological configurations of spin hopfions, torons, and monopoles were never observed in electromagnetic fields before yet.

\section{Results}
\noindent{\textbf{Concept}}

Torons are a special class of spin texture with topological defects localized in real space $\mathbb{R}^3$, which can be described by $\mathbf{s}(x,y,z)=[s_x(x,y,z),s_y(x,y,z),s_z(x,y,z)]$ ($s_x^2+s_y^2+s_z^2=1$), as depicted in Fig.~\ref{f1}\textbf{a}. It includes two singularities where spin vanishes on axis, at positions noted as $z=\pm z_\text{m}$ without loss of generality, as highlighted by the red and blue dots in Fig.~\ref{f1}\textbf{a}, and the main particle-like topological structure is localized within the region between the pair of monopoles, i.e. the regions outside are of spin up background. Each spin singularity is a monopole-like defect, and all orientations of spin are covered exactly once on the sphere around the defect. Such defect is classified by the homotopy group $\pi_2(S^2)=\mathbb{Z}$. The monopole textures are not conventional hedgehog type (source or sink) but hyperbolic type with helicity of transverse spin component, {see inserts cross-section saddle textures of the hyperbolic monopole in Figs.~\ref{f1}\textbf{a2} and \ref{f1}\textbf{a3}}.
%, also usually called Bloch points in ferromagnet~\cite{im2019dynamics} \YS{(correct or not? Bloch point is just helicity?)}. 
The two monopoles also have opposite polarity, i.e. one with spin up/down pointing inward and the other outward, configuration of which is also called monopole-antimonopole pair~\cite{wang2022formation}. 
%\YS{(How to strictly define the $+$ and $-$?)} 
The main particle-like configuration of toron is embedded between the monopoles, where a series of 2D skyrmion textures locate at the intermediate transverse planes, $z\in(-z_\text{m},z_\text{m})$. Such a texture can be classified by the homotopy group $\pi_2(S^2)=\mathbb{Z}$ using a stereographic projection from $S^2$ to $\mathbb{R}^2$ ($S^2\backslash\left\{ 0 \right\}\cong\mathbb{R}^2$). The skyrmions can possess continually changing helicity along $z$ so as to induce chirality of the whole toron, while the topological number maintains unity ($|N_\text{sk}|=1$). The Skyrme number is defined by $N_\text{sk}=\frac{1}{4\pi}\iint B_z^{(\text{em})}\text{d}x\text{d}y$, in which the emergent magnetic field $\mathbf{B}_\text{em}=(B_x^{(\text{em})},B_y^{(\text{em})},B_z^{(\text{em})})$ is defined as $B_i^{(\text{em})}=\varepsilon_{ijk}\mathbf{s}\cdot(\partial_j\mathbf{s}\times\partial_k\mathbf{s})/2$, $\{i,j,k\}\equiv\{x,y,z\}$, and $\varepsilon_{ijk}$ is the Levi-Civita tensor, see \textbf{Supplementary Material S2} for detailed calculation method. The skyrmions terminate at the monopoles ($z=\pm z_\text{m}$) and transits into the trivial uniform background ($N_\text{sk}=0$).

The chiral toron spin texture can be vividly revealed by its fibration representation, i.e. the isospin line cluster, see Fig.~\ref{f1}\textbf{b}. The isospin fibers connect the two hyperbolic spin monopoles torsionally with layer-by-layer structure, from the inner to outer corresponding to the $s_z$ component from $-1$ to $1$, thus the whole configuration forms a fiber bundle with chirality and each layer resembles the pinwheel structure (Fig.~\ref{f1}\textbf{c}). 

A toron also includes sink and source type monopoles in its emergent magnetic field $\mathbf{B}_\text{em}$ at the same positions of hyperbolic spin monopoles, see Fig.~\ref{f1}\textbf{d}. $\mathbf{B}_\text{em}$ streamlines from the
monopole to the antimonopole, which act as source and sink of the field, exactly follow the route of toron fibration, i.e., the isospin lines, see \textbf{Supplementary Material S3} for a mathematical proof. 
%This effect can also be reminiscent of the Dirac string in Dirac monopole~\cite{ray2014observation}. In Dirac monopoles, the Dirac string is an infinitesimally thin solenoid attached to the monopole with quantized magnetic flux. Therefore, the emergence of a skyrmion with nontrivial $\pi_2(S^2)$ field topology in a trivial field background is mediated by the $\pi_2(S^2)$ defects, which are a pair of monopoles in $\mathbf{B}_\text{em}$.
The emergent magnetic field can also be reminiscent of the Dirac monopole~\cite{ray2014observation}. In Dirac monopoles, except the infinitesimally thin Dirac string, the magnetic flux emanates from the monopole to the anti-monopole. Here the emergence of a skyrmion texture with nontrivial $\pi_2(S^2)$ field topology in a trivial field background, exhibiting defects also classified by $\pi_2(S^2)$, leads to the quantization of such magnetic flux and monopole charges. \\[4pt]

\begin{figure*}
\centering
  \includegraphics[width=\linewidth]{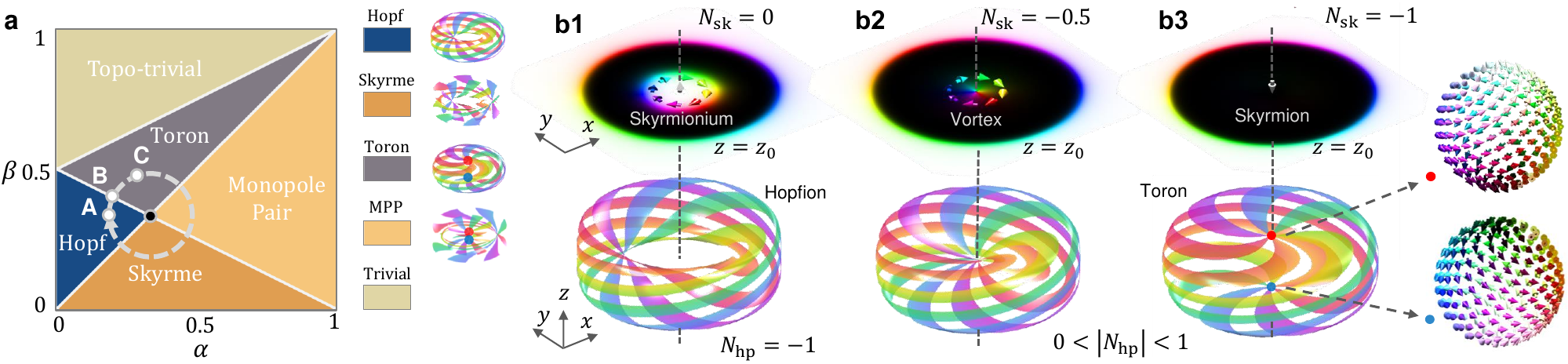}
  \caption{\textbf{Topological phase transition of photonic spin toron.} \textbf{a}, Phase diagram of all topological nontrivial and trivial states versus coefficients $\alpha$ and $\beta$. To visualize the transitions, a loop along the circular path highlights various topological states, and corresponding transformations between the hopfion (Hopf), toron, monopole pair (MPP) and skyrmionium (Skyrme) states are displayed (see \textbf{Supplementary Video} for a dynamic evolution of these transitions). \textbf{b}, There example states in the evolution from a photonic spin hopfion to a toron, with three panels (\textbf{b1–b3}) of corresponding spin texture at central transverse plane (top, with Skyrme number marked) and fibration representation (bottom, with Hopf number marked), respectively, corresponding to three states A, B, and C marked in panel \textbf{a}. In calculation of $N_\text{sk}$ and $N_\text{hp}$, the topological number density integral needs to avoid the neighbourhood of singularity, see \textbf{Supplementary Material S2} for details.}
  \label{f2}
\end{figure*}

\noindent\textbf{Creation of photonic spin torons}

Torons were only previously studied in condensed matter systems such as liquid crystals and magnets, but have never been studied in photonic fields.
%The nonpolar simplified scheme of toron was observed in liquid crystal systems before, however the exact polar torons have never been demonstrated in any physical system. 
Although hopfion textures were created in optical fields using the Stokes vector~\cite{sugic2021particle,shen2023topological}, torons cannot be created using the Stokes vectors, since the polarization field $[E_x, E_y]^T$ cannot form point-like defects in 3D space. We provide rigorous proof in the \textbf{Supplementary Material S7}. However, the photonic spin density allows the existence of point-like defects in 3D space. Hereinafter, we show that the exact topological structures of polar torons can be constructed by photonic spin in a structured light field.

%\YS{To Haiwen: We need state the challenge, we need to say it is impossible to be made by field vector, or Stokes vector. We can break the limit, make the impossible into the possible in spin, we did. Not sure if at here.}

We exploit the photonic spin or the SAM density vector. The total SAM density vector of a monochromatic electromagnetic field in real 3D space is defined as the sum of both electric and magnetic spin components:
\be
{\bf S} = \frac{1}{4\omega}\left[\epsilon_0\mathrm{Im}({\bfE}^*\times{\bfE}) + {\mu_0}\mathrm{Im}({\bf H}^*\times{\bf H})\right],
\label{spin}
\ee
where $\bfE$ and $\bf H$ are the complex vectors of the electric and magnetic fields, respectively, $\epsilon_0$ is the vacuum permittivity, and $\mu_0$ is the vacuum permeability, and $\omega$ is the angular frequency of the light. Notably, we use normalized spin $\bf s=\bf S/|\bf S|$ to construct topological spin textures.

\begin{figure*}
\centering
  \includegraphics[width=\linewidth]{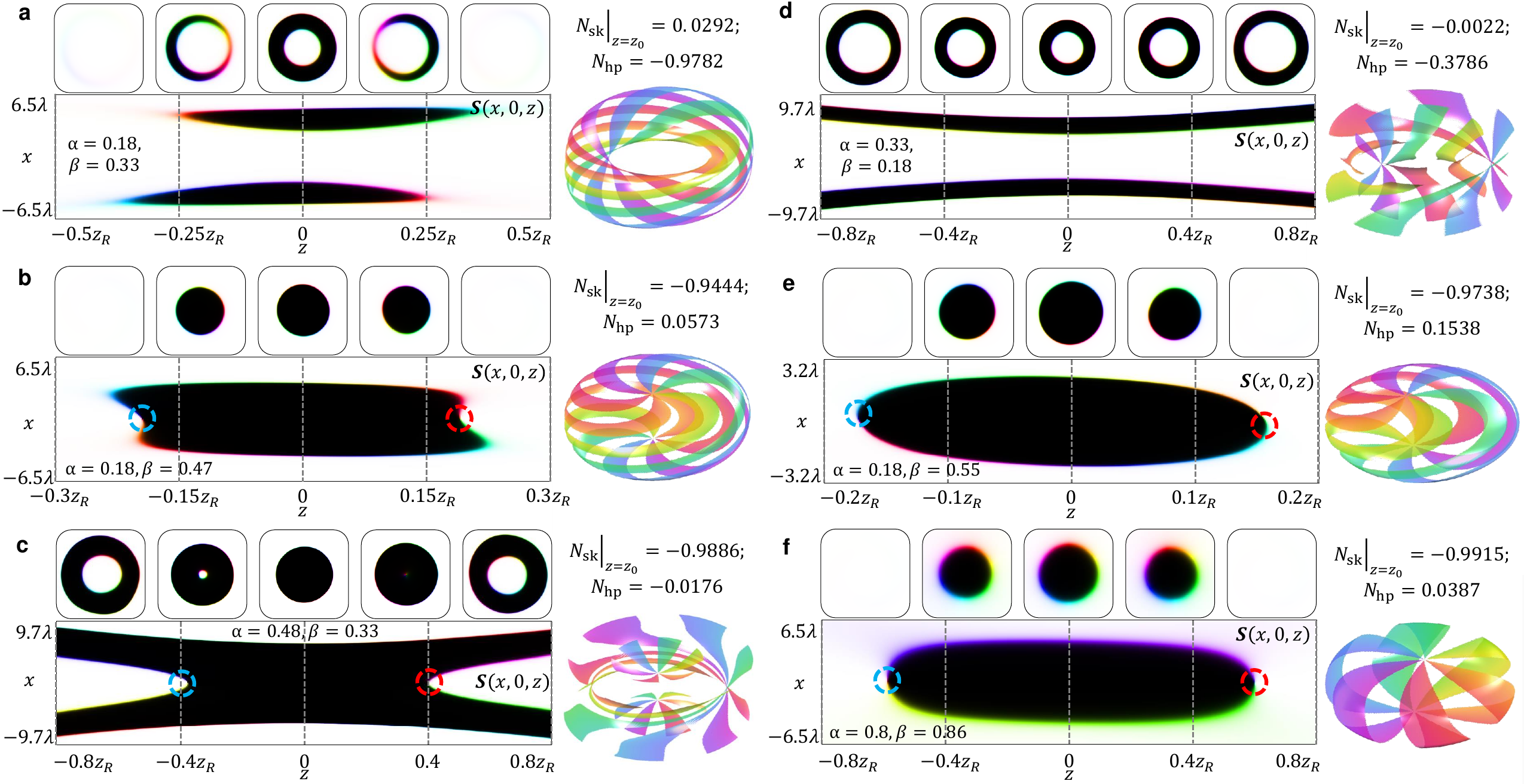}
  \caption{\textbf{Experimental results of photonic spin hopfion, toron, skyrmionium, and monopole pair.} \textbf{a-h}, The propagation tomographies of optical spin textures for various typical topological states for a hopfion (\textbf{a}), toron (\textbf{b}), monopole pair (\textbf{c}), skyrmionium (\textbf{d}), and other two torons with different helicity (\textbf{e}) and chirality (\textbf{f}), with corresponding values of $\alpha$ and $\beta$ marked, including distributions of longitudinal $x$-$z$ cross-section and selective transverse $x$-$y$ cross-sections (each panel left) and alongside the corresponding assemblies of isospin line clusters (each panel right). The Skyrme number at central transverse plane ($N_\text{sk}|_{z=z_0}$) and Hopf number ($N_\text{hp}$) for each state are indicated in corresponding panel.}
  \label{f3}
\end{figure*}

To construct spin torons, both transverse and longitudinal components of electromagnetic field should be considered according to Eq.~(\ref{spin}), given by $\bfE=E_x\bfe_x+E_y\bfe_y+E_z\bfe_z=\bfE_\perp+E_z\bfe_z$ and $\bfH=H_x\bfe_x+H_y\bfe_y+H_z\bfe_z=\bfH_\perp+H_z\bfe_z$, and we design the structured transverse electric field as:
\be
 {\bfE}_\perp(\bfr)=\alpha\psi_{0,0}(\bfr)\bfeR+\left[\beta\psi_{0,0}(\bfr)-(1-\beta)\psi_{0,1}(\bfr)\right]\bfeL,
\label{lg}
\ee
where $\bfr=(x,y,z)$ is spatial coordinate frame, $\bfe_x$, $\bfe_y$, and $\bfe_z$ are  Cartesian unit vectors, $\bfeR$ and $\bfeL$ are unit vectors of right-circular polarization (RCP) and left-circular polarization (LCP) components of transverse electric field transformable from $\bfe_x$, $\bfe_y$, $\psi_{\ell,p}$ is Laguerre-Gaussian (LG) mode with azimuthal and radial indices, $\ell$ and $p$, respectively (see \textbf{Supplementary Material S1} for expression), and $\alpha$ and $\beta$ are complex amplitude coefficients. According to the slowly varying envelope approximation, the transverse magnetic field can be expressed as a function of the electric counterparts: 
\be
 {\bfH}_\perp(\bfr)=H_x\bfe_x+H_y\bfe_y=-\frac{E_y}{Z_0}\bfe_x+\frac{E_x}{Z_0}\bfe_y,
\ee
where $Z_0$ is the vacuum impedance. The longitudinal components of the fields are obtained, using Gauss's law ($\nabla\cdot\bfE=0$, $\nabla\cdot\bfH=0$), with the paraxial approximation given by Lax~\cite{Lax_method}, as $E_{z} =(i/k)\nabla\cdot{\bfE}_\perp(\bfr)$ and $H_{z} =(i/k)\nabla\cdot{\bfH}_\perp(\bfr)$.
%\be
%E_{z} = \frac{i}{k}\left( {\frac{\partial E_{x}}{\partial x} + \frac{\partial E_{y}}{\partial y}} \right), H_{z} = \frac{i}{k}\left( {\frac{\partial H_{x}}{\partial x} + \frac{\partial H_{y}}{\partial y}} \right).
%\ee
Therefore, the spatial distribution of the three components of the electromagnetic fields, $(E_x,E_y,E_z)$ and $(H_x,H_y,H_z)$, and the resultant spatial spin distribution ${\bf S}({\bf r})$ can be obtained. 
 
By tuning the value of $(\alpha,\beta)$, diverse topological phases can be observed in the spin distribution. Although $(\alpha,\beta)$ can be complex numbers, here we restrict them to be real in the range $(0,1)$ due to the following reasons. First, the intermodal phase between the LCP and RCP envelope does not affect the spin texture, see \textbf{Supplementary Material S4} for a proof. Second, although the phase of $\beta$ changes the spin texture, we choose $\beta$ to be real to obtain a texture centered at the focal plane. 
Five distinct topological phases are classified in the real $(\alpha,\beta)$ map, see Fig.~\ref{f2}\textbf{a}:
\begin{itemize}
\setlength{\itemsep}{0pt}
\setlength{\parsep}{0pt}
\setlength{\parskip}{2pt}
\item Hopfion: For $\beta>\alpha$ and $2\beta<1-\alpha$, the spin pattern exhibits smooth spatial texture with hopfion topology, with isospin contours as layer-by-layer torus knots fulfilling Hopf fibration. A spin skyrmionium structure is located at the focal plane (Skyrme number $N_\text{sk}=1+(-1)=0$), while the texture gradually evolves into a constant field towards the infinity. See an example in Fig.~\ref{f2}\textbf{b1}(Hopf number $N_\text{hp}=-1$). 
    
\item Toron: For $\beta>\alpha$ and $1+\alpha>2\beta>1-\alpha$, the spin pattern exhibits toron topology. During the transition from hopfion to toron, the inner skyrmion or the torus hole decreases, transforming the skyrmion into a vortex singularity ($N_\text{sk}=-0.5$). Eventually, the vortex texture vanishes, revealing the rest as a 2D skyrmion texture ($N_\text{sk}=-1$), indicating the emergence of a spin toron with two monopoles. %Moreover, as the state departs from the critical state, the absolute value of the Hopf number $|N_\text{hp}|$ rapidly decrease below 1, indicating the destroying of hopfion topology.
    
\item Monopole pair (MPP): For $\beta<\alpha$ and $2\beta>1-\alpha$, the spin pattern still possess two spin monopoles (monopole-antimonopole pair) but without toron topology. During the transition from toron to MPP, the connection of isospin lines between the two monopoles is broken, in other words, the emergent magnetic streamlines are opened and splayed to infinity.
    
\item Skyrmionium (Skyrme): For $\beta<\alpha$ and $2\beta<1-\alpha$, the spin texture shows only skyrmionium topology at arbitrary transverse plane at any \textit{z} position, we call it photonic spin skyrmionium beam, which is in contrast to other cases that the skyrmionium can only be observed in transverse plane in a limited range of \textbf{z}. During the transition from MPP to skyrmionium, a pair of spin monopole and antimonopole emerge and annihilate each other.
\item Trivial phase: For $2\beta>1+\alpha$, the field is of a nearly uniform texture.
\end{itemize}

In the topological phase diagram, there is a four-phase junction, with the black solid dot in Fig.~\ref{f2}\textbf{a}, indicating the continuous transition between arbitrary two phases of the four nontrivial states.

We note that all of these topological optical spin structures have not been experimentally observed before. Exploiting advanced structured light technology, see \textbf{Methods}, here we can experimentally generate all of these topological nontrivial optical spin states and control their shape and chirality. Figure.~\ref{f3}\textbf{a-d} show our experimental results of the four typical topological nontrivial states, i.e. photonic spin hopfion, toron, MMP, and skyrmionium. In the toron state, the helicity and chirality are two unique properties. Here, the helicity determines how strongly twisted the isospin lines are between the two spin monopoles, while the chirality defines whether the isospin fiber twists are right- or left-handed. We can also control the different helicity and chirality in our experiment. Figure~\ref{f3}\textbf{e} shows a toron whose helicity is weaker than that of Fig.~\ref{f3}\textbf{b}, while Fig.~\ref{f3}\textbf{f} shows a toron with opposite chirality to the cases of Figs.~\ref{f3}\textbf{b} and \ref{f3}\textbf{e}. \\[4pt]

\begin{figure*}
\centering
  \includegraphics[width=0.95\linewidth]{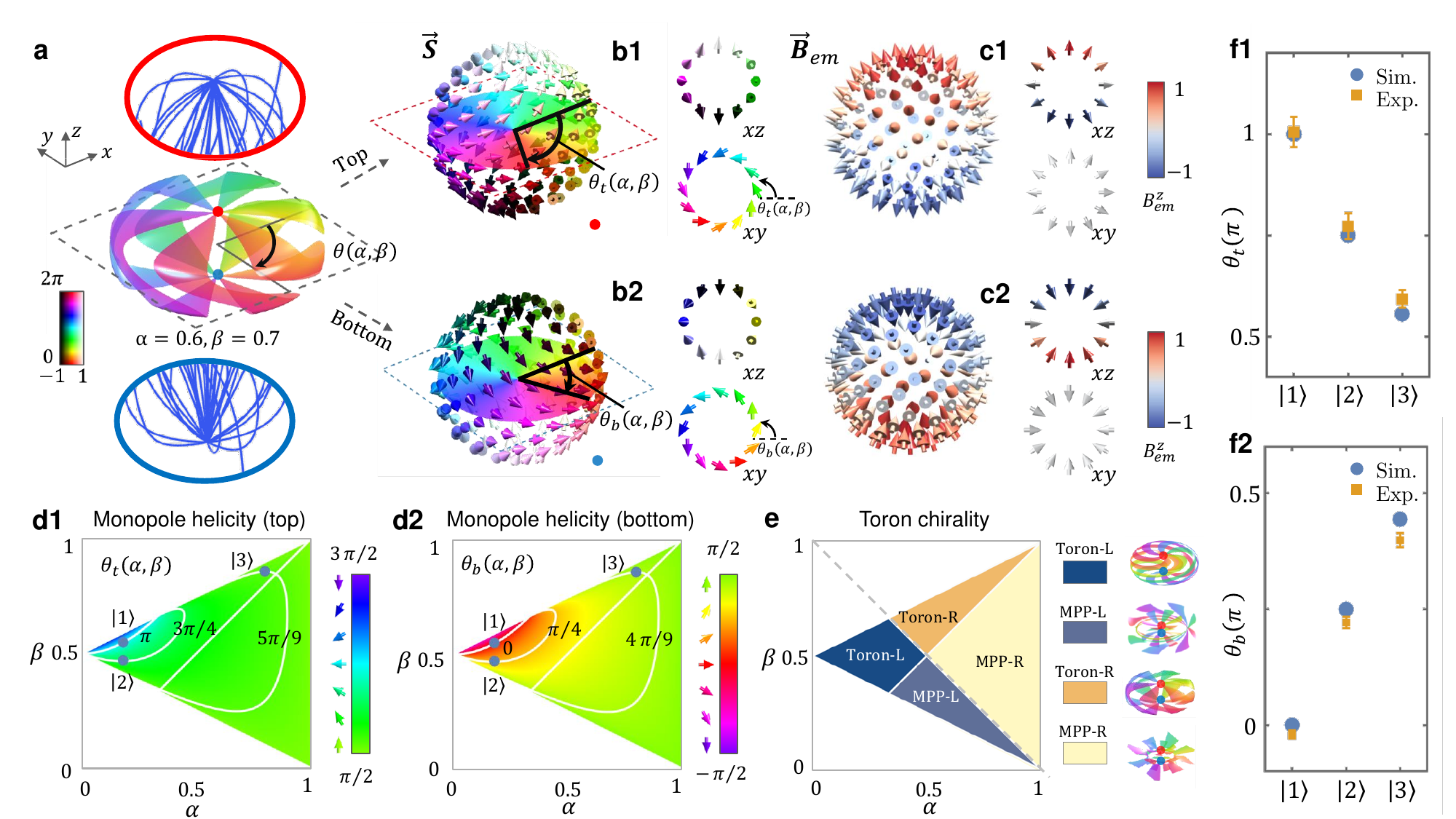}
  \caption{\textbf{Control of monopole helicity and toron chirality.} \textbf{a}, Highlights of the two monopole defects in a toron coexisting in both the spin texture and the emergent magnetic field. \textbf{b1-b2}, The spin textures around the top (\textbf{b1}) and the bottom (\textbf{b2}) monopoles, and the winding on their $xz$ and $xy$ cross-sections. The helicity of the spin monopole is defined by the transverse spin component, exhibiting a vortex structure. This can be described by $\psi_m=\varphi+\theta_\text{t,b}(\alpha,\beta)$, where $\varphi$ is the azimuthal angle of the monopole's transverse plane, and $\theta_\text{t,b}(\alpha,\beta)$ are additional angles controlled by the parameters $\alpha$ and $\beta$ for the top and bottom monopoles, enabling the controlled tuning of monopole helicity. \textbf{c1-c2}, The emergent magnetic field textures around the top (\textbf{b1}) and the bottom (\textbf{b2}) momopoles, and the winding on their $xz$ and $xy$ cross-sections. For the monopole of the emergent magnetic field, only source and sink types exist, regardless of the helicity of the spin monopole. \textbf{d1-d2}, Phase diagrams of helicity of the top (\textbf{d1}) and the bottom (\textbf{d2}) monopoles. The tunable ranges of $\theta_\text{t,b}(\alpha,\beta)$ are shown with contour lines, where the top monopole helicity $\theta_\text{t}$ is tunable from $\pi⁄2$ to $3\pi⁄2$, while bottom monopole $\theta_\text{b}\in(-\pi/2,\pi/2)$, resulting in hyperbolic monopoles generated in this study. \textbf{e}, Phase diagram illustrating the chirality of the toron, with distinct regions for left- and right-handed monopole pair (MPP) and toron states. \textbf{f1-f2}, Experimentally obtained tunable angles $\theta_\text{t}$ and $\theta_\text{b}$ for three states $\ket{1}$, $\ket{2}$ and $\ket{3}$ marked in (\textbf{d1-d2}), align well with theoretical simulations. }
  \label{f4}
\end{figure*}

%\begin{figure*}
%\centering
%  \includegraphics[width=0.95\linewidth]{F5.pdf}
%  \caption{\textbf{Experimental tuning of spin monopole helicity.} The 3D top (a1-c1) and bottom (a2-c2) monopole textures of three torons (denoted as $\ket{1}$, $\ket{2}$ and $\ket{3}$) are shown for different parameters $\alpha$ and $\beta$, with each accompanied by their corresponding 2D vortex structures and one-layer 3D transverse spin density distributions. These panels illustrate the helicity transformations between the different states. (d1-d2) present the experimentally obtained tuning angles $\theta_t$ and $\theta_b$ for the top and bottom monopoles, alongside theoretical reference values, showing good agreement between simulation and experimental results. And the insets show clearly the positions of the three states on the tunable angle map.}
%  \label{f5}
%\end{figure*}

\noindent\textbf{Tuning spin monopole helicity}

Synthetic magnetic monopole is an important physical concept, however, how to tune the monopole texture still remains elusive to study in any physical system. 
%Here we fill this gap. In the preceding sections, we demonstrated the topological phase transition of nontrivial states and examined the tuning of helicity and chirality in torons via the coefficients . 
In this section, we will fill this gap by further studying the physical mechanism of the helicity and chirality tunability through the coefficients $(\alpha,\beta)$.

As depicted in Fig.~\ref{f4}\textbf{a}, the monopole texture appears in both the spin texture and the emergent magnetic field of torons or MPPs. The vortex-like spin orientation $\psi_m$ at the central transverse cross-sections, $z=\pm z_m$, where $s_z=0$, is used to describe the helicity of monopole defects. Adjusting $(\alpha,\beta)$ alters the helicity angle $\theta_{t,b}(\alpha,\beta)$ for the top and bottom monopoles, changing the transverse spin orientations while maintaining two polar points (spin up/down) unchanged. Despite the variation in the spin texture helicity, the monopole of the emergent magnetic field, source (top) and sink (bottom), remains unchanged.

In Figs.~\ref{f4}\textbf{d1} and ~\ref{f4}\textbf{d2}, the tuning ranges of $\theta_{t,b}(\alpha,\beta)$ for torons and MPPs are highlighted within triangular areas, excluding the Hopf and the trivial regions. The ranges are complementary $\theta_{t}(\alpha,\beta)+\theta_{b}(\alpha,\beta)=\pi$, dictating helicity and chirality of toron and MPP. However, clarifying helicity and chirality also requires considering the skyrmion texture helicity at $z=z_0$, influenced by $\theta(\alpha,\beta)$. At the $z_0$ plane, Eq.~\ref{lg} shows that RCP and LCP mode components are real functions, resulting in $s_r=0$. Here, we note $(s_r,s_\phi)$ as radial and azimuthal components of transverse optical spin $(s_x,s_y)$. Consequently, the sign of $s_\phi$ depends on the relative intensity rate of RCP and LCP mode component, making the helicity of the skyrmion solely influenced by $s_\phi$. For $\beta<1-\alpha$, leading to $s_\phi<0$ and $\theta(\alpha,\beta)=3\pi⁄2$; conversely, $\beta>1-\alpha$, leading to $s_\phi>0$, $\theta(\alpha,\beta)=\pi⁄2$. Thus, skyrmion textures manifest as only two Bloch types based on $(\alpha,\beta)$, see \textbf{Supplementary Material S4} for detailed explanation.

Utilizing $\theta_{t}(\alpha,\beta)$ and $\theta(\alpha,\beta)$, we can precisely describe the winding of isospin fiber bundles and analyze toron helicity and chirality. The bundles radiate from the top monopole defect ($z=z_m$), then deviate and pass through the $z_0$ plane at an azimuthal angle that is different from the origin of the fiber by $\theta_H(\alpha,\beta)=\theta_{t}(\alpha,\beta)-\theta(\alpha,\beta)$. Positive $\theta_H(\alpha,\beta)$ indicates right-handed chirality, while negative values indicate left-handed chirality. Although isospin fibers break in MPPs, the rotation trend persists, allowing $\theta_H(\alpha,\beta)$ to dictate both toron and MPP chirality. The corresponding chirality phase diagram is shown in Fig.~\ref{f4}\textbf{e}. %Helicity magnitude, shown in the Fig.~\ref{f4}\textbf{e} inset, correlates with $\theta_H(\alpha,\beta)$, where higher values denote stronger helicity.
Helicity is linked to the magnitude of 
$\theta_H(\alpha,\beta)$, with higher values indicating stronger helicity, see \textbf{Supplementary Material S8} for more details.

Finally, in Figs.~\ref{f4}\textbf{f1} and \textbf{f2}, we present three groups of tunable angles labeled as $\ket{1}$, $\ket{2}$, $\ket{3}$, observed from the experimental toron structures in Figs.~\ref{f3}\textbf{b}, ~\ref{f3}\textbf{e}, and ~\ref{f3}\textbf{f}. The tunable angles $\theta_{t,b}(\alpha,\beta)$ for these examples show good agreement with simulation predictions (see \textbf{Supplementary Materials S3} for detailed comparison of experimental and simulated spin monopole textures).
 \\[4pt]

\noindent\textbf{Discussion}

In conclusion, we proposed the first physical model to generate torons in optical fields, with polar spin chiral quasiparticles with both topological textures and monopole defects. We experimentally generated and observed the torons in photonic spin distributions of focused vectorial structured light. We also demonstrated the topological phase transition in our topologically structured spin field, including nontrivial topological phases of toron, hopfion, skyrmionium, and monopole pair both theoretically and experimentally. We argue that each nontrivial topological phase constitutes, to the best of our knowledge, the first experimental report of free-space optical spin textures. We also demonstrated the chirality control of torons and helicity control of monopoles (including the monopoles in both toron phase and monopole pair phase), where all the hedgehog (sink and source) type and helical (hyperbolic) type Bloch points can be realized in this same system. Our method constitutes a platform for the control of 2D and 3D topological phases, which we offers greater flexibility compared to other synthetic magnetic monopole counterparts. 

Compared with the previous generation of torons and monopoles in condensed matter systems~\cite{castelnovo2008magnetic, ray2014observation, ackerman2017diversity, kanazawa2020direct}, optical torons and monopoles offer more controllable degrees of freedom and versatile tunability. Optical spin topological structures with transverse spin angular momentum also lead to nontrivial light-matter interactions~\cite{bliokh2015spin,aiello2015transverse}. Applications of using such optical spin textures in controlling the on-demand orientation or rotation of nanoparticles and transferring the topology of wave into matter can be envisioned.

In contrast to prior liquid-crystal based nonpolar torons~\cite{smalyukh2010three,ackerman2017diversity,tai2020surface,peixoto2024mechanical,zhao2023liquid,zhao2023topological,poy2022interaction,tai2024field}, our photonic spin torons are of exact polar scheme. Due to the nonpolar head-tail symmetry of liquid crystal, the prior toron models can only simulate spin texture after artificial vectorization, thus, for skyrmion topology, the emulated order-parameter space and homotopy group are $S^2/\mathbb{Z}_2$ and $\pi_2(S^2/\mathbb{Z}_2)$, respectively. However, photonic spin is a natural polar vector, the skyrmion topology is classified exactly using the homotopy group $\pi_2(S^2)$. In contrast to prior topological quasiparticles realized in optical polarization Stokes vector fields, including Stokes skyrmions~\cite{gao2020paraxial,shen2021generation} and hopfions~\cite{sugic2021particle,shen2023topological},  the optical spin quasiparticles may present higher topological stability, as the spin texture is not perturbed by intermodal phases. Such creation and control of photonic spin torons and monopoles will also stimulate their observations in other physical wave systems and condensed matter, such as acoustic, elastic, water waves and chiral magnets.

Finally, spin torons are created for the first time in optical systems, together with the first-time experimental observations of spin skyrmions, hopfions, and monopoles in free-space structured light. This may open new platforms for topologically protected and higher-dimensional information encoding and transfer for advanced communication technologies and topologically nontrivial light-matter interactions.

\section{Methods}
\noindent\textbf{Generation and detection of optical spin texture}

The transverse mode ${\bfE}_{\perp}({\bf r})$ was generated using a self-stabilizing polarization Mach-Zehnder interferometer with two beam displacers (see \textbf{Supplementary Material S5} for detailed experimental setup). To observe the photonic spin texture ${\bf S}({\bf r})$, the wavefunction of generated spatial mode ${\bfE}({\bf r})$ must be determined. Here, A ``complex-amplitude profiler'' was employed to capture the full spatial wavefunction of ${\bfE}_{\perp}({\bf r})$, first. The RCP and LCP components of ${\bfE}_{\perp}({\bf r})$ were combined with an ancillary beam from the same laser source via a polarizing beam splitter (PBS) to generate a vector mode, respectively. Spatially resolved Stokes parameters, ${S}_0({\bf r})$, ${S}_1({\bf r})$, ${S}_2({\bf r})$, and ${S}_3({\bf r})$, enabled the reconstruction of the full spatial wavefunction of the field ${\bfE}_{\perp}({\bf r})$ at the Fourier plane (see \textbf{Supplementary Materials S5} for details on experimental setup and theoretical principles). The longitudinal field ${E}_{z}({\bf r})$  was then obtained through Gauss's law.\\[4pt]

\noindent\textbf{Digital propagation for 3D topological structures}

Since the photonic spin textures observed form 3D topological structures along the propagation direction $z$, a polarization-insensitive digital micromirror device (DMD) was employed for digital propagation on both $E_R$ and $E_L$ components of ${\bfE}_{\perp}({\bf r})$. The phase corresponding to the desired propagation distance was encoded in the DMD using binary holograms. This technique, combined with the ``complex-amplitude profiler'', enabled the camera, located at the back focal plane of the Fourier lens (L3), to capture the 3D spatial wavefunction at various distances without any mechanical movement along the $z$-axis. Additionally, due to the beam radius of $8\lambda$ being too small for the CMOS camera (with a pixel size of 3.45 $\upmu$m) to capture with sufficient resolution, an interpolation method was applied to enhance image quality (refer to the \textbf{Supplementary Material S6} for details).
 \\[4pt]
\bibliographystyle{naturemag}

\textbf{Acknowledgments.}
Y. Shen acknowledges the support from Nanyang Technological University Start Up Grant, Singapore Ministry of Education (MOE) AcRF Tier 1 grant (RG157/23), MoE AcRF Tier 1 Thematic grant (RT11/23), and Imperial-Nanyang Technological University Collaboration Fund (INCF-2024-007). 
 \\[4pt]
 \textbf{Author contributions.}
Y. S. conceived the basic idea of the work, H. Wu designed and performed the experiments. H. Wang performed the basic theory, H. Wu, N. M.-C., H. Wang and Y. S. analyzed the experimental data and wrote the draft. All authors took part in discussions, interpretations of the results, and revisions of the manuscript, Z. Z., C.-W. Q. and Y. S. supervised the project.
 \\[4pt]
\textbf{Competing interests.}
The authors declare no competing interests.
 \\[4pt]
 \textbf{Data and materials availability.}
All data needed to evaluate the conclusions in the paper are
present in the paper and the Supplementary Materials.
 \\[4pt]
%\\[8pt]
%\noindent
%\textbf{Author contributions}\\
%Y.S. conceived the idea and wrote the paper, C.H. designed the GRIN lens cascades, performed simulation and experiment of the GRIN lens based skyrmionic beam generation, with the input of Z.S., B.C., H.H., Y.M., J.A.J.F., S.J.E., S.M.M., and M.J.B. together. Y.S, C.H., S.M.M., and A. F. contributed to data analysis and manuscript revision.
%\\[8pt]
%\noindent
%\textbf{Competing interests}\\
%The authors declare no competing interests.

\bibliography{sample}

\end{document}